\documentclass[aip,jcp,%,superscriptaddress
,Ndraft
,showpacs
,preprint
]{revtex4-1}
\usepackage{graphicx}% Include figure files
\usepackage{amsmath}% for double integral sign
\usepackage{dcolumn}% Align table columns on decimal point
\usepackage{multirow}
\usepackage{amssymb}
\usepackage{isomath}
\usepackage{rotating}
\usepackage[usenames]{color}
\usepackage{epsf}
\usepackage{ifthen}
\usepackage{subfigure}
\usepackage{psfrag}
\usepackage{float}
\usepackage{subeqnarray}

\newcommand{\beq}{\begin{eqnarray}}
\newcommand{\eeq}{\end{eqnarray}}
\begin{document}
\title{Electronic properties of graphene nano-flakes:\\ Energy gap, permanent dipole, termination effect and Raman spectroscopy}
\author{Sandeep K. Singh}
\email[]{SandeepKumar.Singh@uantwerpen.be}
\author{M. Neek-Amal}
\email[]{ neekamal@srttu.edu}
\author{F.M. Peeters}
\email[]{Francois.Peeters@uantwerpen.be}
\affiliation{Department of Physics, University of  Antwerpen, Groenenborgerlaan 171, B-2020 Antwerpen, Belgium}%

\date{\today}
\pagenumbering{arabic}

%Abstract

\begin{abstract}

The electronic properties of graphene nano-flakes (GNFs)  with
different edge passivation is investigated by using density functional
theory. Passivation with F and H atoms are considered:
C$_{N_c}$ X$_{N_x}$ (X=F or H). We studied GNFs
 with  $10<N_c<56$ and limit ourselves to the lowest energy configurations.
 % for planar flakes.
  We found that: i) the energy difference $\Delta$ between the highest occupied molecular
orbital (HOMO) and the lowest unoccupied molecular orbital (LUMO)
decreases with $N_c$, ii) topological defects (pentagon and
heptagon) break the symmetry of the GNFs and enhance the electric
polarization, iii) the mutual interaction of bilayer GNFs can be
understood by dipole-dipole interaction which were found sensitive
to the relative orientation of the GNFs,  iv) the permanent dipoles
depend on the edge terminated atom, while the energy gap is
independent of it, and v) the presence of heptagon and pentagon
defects in the GNFs results in the largest difference between the
energy of the spin-up and spin-down electrons which is larger for
the H-passivated GNFs as compared to F-passivated GNFs. Our study
shows clearly the effect of geometry, size,  termination and bilayer
on the electronic properties of small GNFs.This study reveals
important features of graphene nano-flakes which can be detected
using Raman spectroscopy.
\end{abstract}

\pacs{73.22.-f;71.15.Mb}

\maketitle
\section{Introduction}
Graphene nano flakes (GNFs) and graphene nano ribbons (GNRs)
are promising graphene based materials with a size controllable energy band gap,
which may be useful for different technological
applications~\cite{abergel,castroneto}. In particular, these graphene nano-flakes
are important due to their potential for bottom-up fabrication of
molecular devices, spintronics and quantum dot
technology~\cite{son}. Bottom-up and top-down approaches are two
alternatives for the production of GNFs. In the first approach, large
aromatic hydrocarbons are produced using a large variety of chemical
reactions between small molecular units~\cite{Wu,Zhi}. The
top-down method starts with a large piece of graphene sheet and
cuts the GNFs out of it. Single graphene sheets  can be obtained by a variety of
methods, e.g. micromechanical cleavable of a graphite single crystal
~\cite{Geim2004}, starting from graphite oxide~\cite{park2006} or by
chemically unzipping of carbon nanotubes ~\cite{Terrones2009}.

Tight-binding (TB) approximation based on the $\pi$ orbitals of
carbon, the free massless particle Dirac's equation and ab-initio
calculations are three common theoretical methods for studying the
electronic and magnetic properties of GNFs and GNRs. Similar to TB
calculations~\cite{Potasz} or solutions of the Dirac
equation,~\cite{mamadPRB} ab-initio calculations also show that the
GNRs have a non-zero direct band gap~\cite{PRL2006}. Already several
studies have appeared on the electronic and magnetic properties of
small GNRs using various methods~\cite{refs}.  However much less
extensive and systematic studies are available  on  various
properties of the GNFs, e.g. the size, edge termination and
polarization are still poorly understood. Using
tight-binding calculation and Hartree-Fock theory the energy gap
dependence of triangular and hexagonal GNFs on size, shape and edge
were  studied by G\"{u}\c{c}l\"{u} et al.~\cite{Guclu2010}. It was found
that triangular GNFs with zigzag edges exhibit optical transitions
in wide spectral ranges, i.e. from teraHertz up to the UV. Because
of the small size of the GNFs, they can be considered as a
zero-dimensional form of graphene which exhibit very different
properties from  GNRs and bulk graphene. They are promising
 for a variety of applications, e.g. electronic and
magnetic devices with various molecular sizes and shapes.
Graphene nano flakes (graphene quantum dots) can be useful
for light absorption relevant for photovoltaics due to their edge structure
and wide spectrum. GNFs are found to possess unique electronic,
magnetic, and optical properties due to their tunable band gap,
e.g., they can be used in solar cells and LED technology
~\cite{nanolett2010}. They have different corners, mixed
zig-zag and arm-chair edges which provide additional degrees of
engineering freedom. The small size of GNFs leads to discrete energy
levels similar to atomic levels in single atoms. The saturation with
different atoms or molecular groups on the zigzag edges of
rectangular GNFs leads to a spin-polarized ground state with a non
zero total magnetic moment, a spin density, and an electronic energy
gap that strongly depends on the atomic group used to passivate the
dangling bonds~\cite{Zheng2008}. First-principles calculations were
also used to investigate the magnetic properties of GNFs with
triangular shape  and fractal structure~\cite{28,Voznyy} (see
Ref.~\cite{ian} for a review).

Only a few studies on the electronic and magnetic properties of
GNFs have been published, however most of those studies are limited to
 triangular or hexagonal shapes of graphene nano flakes with  pure zig-zag or arm-chair edges
  without considering edge passivation,
stability~\cite{siefert}, polarization effects and melting phenomenon~\cite{sandeep}. Here
we report on  various GNFs with different shapes
 that are the most energetically favorable configurations
 for given number of carbon atoms in the flake, as found
in our previous work~\cite{kosi1}. Such nano flakes were found to
have lower melting temperature than graphene~\cite{sandeep}.
There are few works on bilayer GNFs, e.g. G\"{u}\c{c}l\"{u} et al.~\cite{Guclu2011} studied graphene bilayer triangular
quantum dots and found that it exhibits a shell of degenerate states
at the Fermi level. Moreover  applying a vertical electric field
on bilayer GNF can turn off or reduce the total spin to a single
localized spin. In the present study we focus on the electronic properties
and permanent polarization of those GNFs with two different edge
passivation. We show that the electric dipole moments strongly
depend on the symmetry of the GNFs and the type of edge passivation.
Our findings for GNFs with $n$-fold symmetry can be extended
to larger flakes without loss of generality, they are always
un-polarized independent of the type of edge terminated atoms. We
also found that the stability of bilayer GNFs depends on the mutual
orientation of the  permanent dipole moments.

 This paper is organized as follows. In Sec. II we give a short review on
 the most important polarization effects that we need in our study on the  GNFs.
 In Sec. III, we introduce the
 used DFT calculation. Next in Sec. IV and Sec. V
the effects due to symmetry and the energy gap of GNFs and its consequences
on the electrical polarization are presented, respectively. The density
of states and possible half-metallicity in larger GNFs are also
discussed. Sec. VI contains our main results and a discussion on the single sheet GNFs. In
Sec. VII we present the results for typical bilayer GNFs and Sec. VIII presents results that are relevant for
 Raman spectroscopy. Finally, we conclude the paper in Sec. IX.
\begin{figure}
%\begin{center}
\centering
\includegraphics[width=0.9\linewidth]{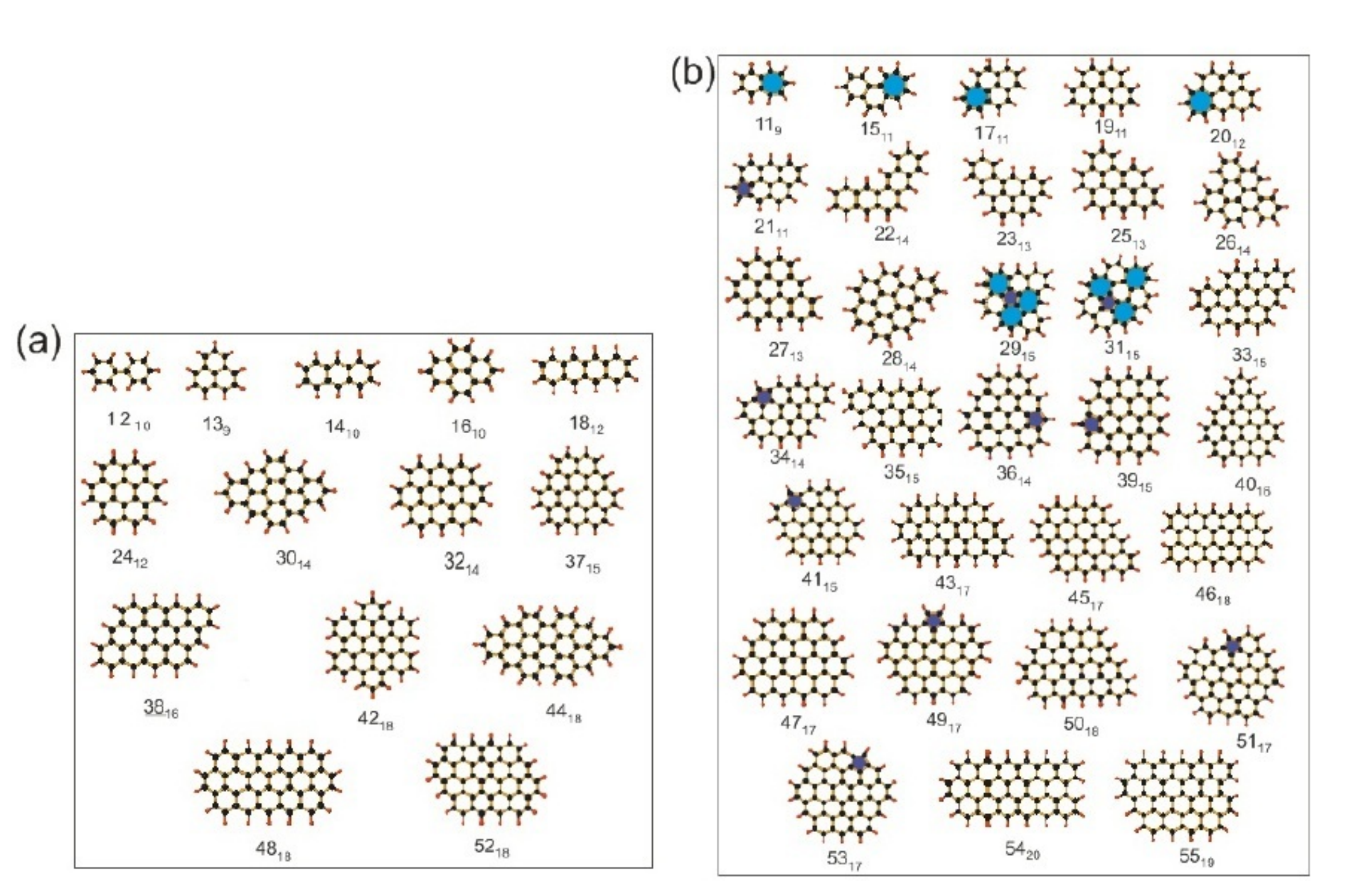}
\caption{(Color online) The minimum energy
configurations of the studied GNFs. The black balls indicate C atoms   and the red balls can be H
or F atoms which saturate the edges. (a) The GNFs with n-fold
symmetries (the underlined GNFs have 2-fold symmetry). (b) GNFs
without n-fold symmetry. The pentagons have indigo color and the
heptagons are in blue. The subindex in each GNFs refers to the
 number of H or F atoms and the main number equals the number of C-atoms in the flakes. The
 shaded polygons are not hexagons.\label{figmodel}}
%\end{center}
\end{figure}

\begin{figure}
%\begin{center}
\centering
\includegraphics[width=0.6\linewidth]{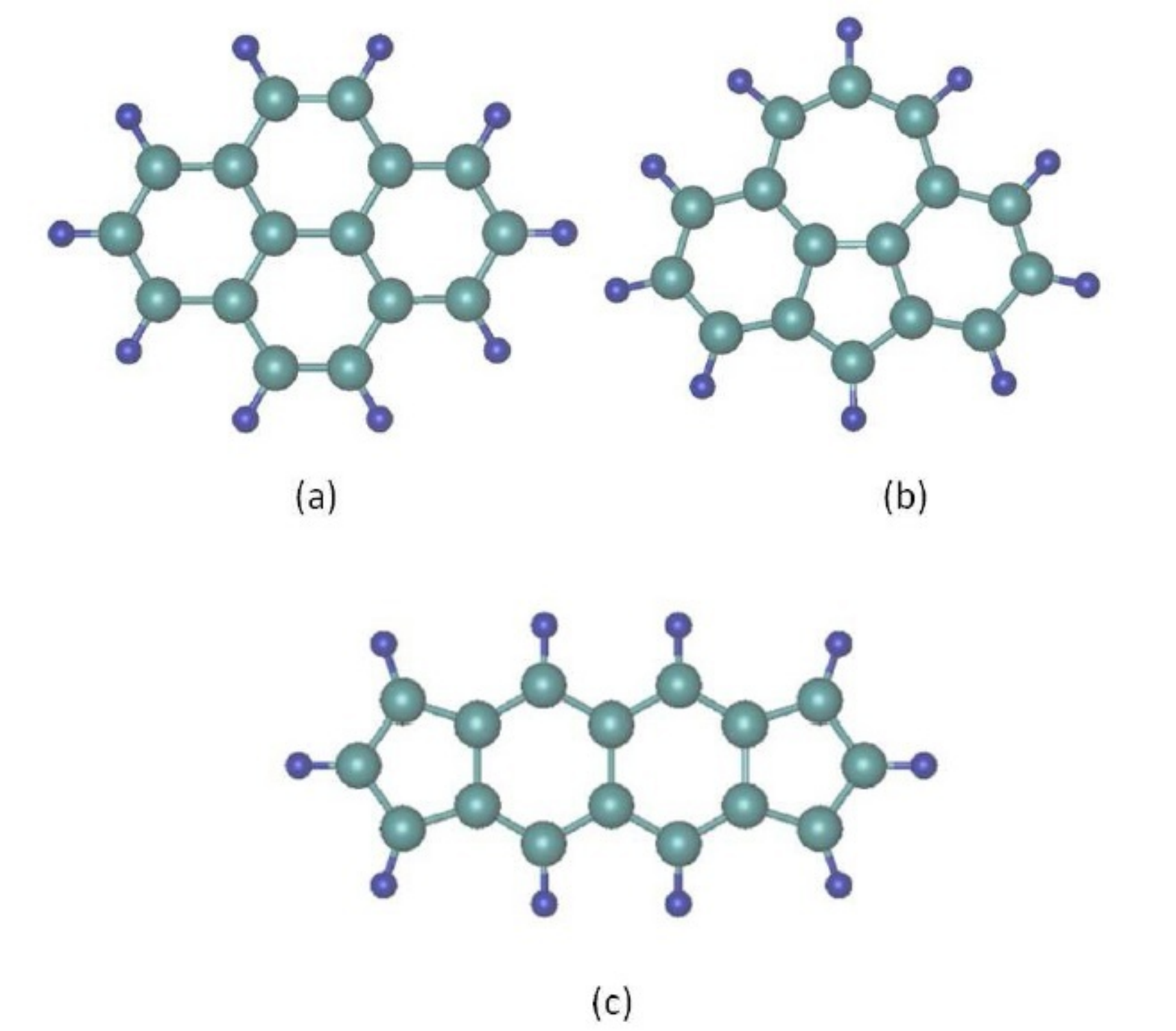}
 \caption{(Color online) Three minimum energy isomers of C$_{16}$X$_{10}$ whose energy are given in Table~I.
  \label{isomers}}
%\end{center}
\end{figure}

\begin{table}[tp]%
\caption{Binding energy for the three isomers of C$_{16}$X$_{10}$
that are depicted in Fig.~\ref{isomers}.}
\begin{tabular}{c  | c  }
\hline
       & Binding energy (eV)                  \\
 \hline

C$_{16}$H$_{10}$(a)  & -170.49                      \\
C$_{16}$H$_{10}$(b)  & -168.10                      \\
C$_{16}$H$_{10}$(c)  & -167.78                      \\
 \hline
C$_{16}$F$_{10}$(a)  & -171.81                      \\
C$_{16}$F$_{10}$(b)  & -169.47                     \\
C$_{16}$F$_{10}$(c)  & -169.62                      \\
 \hline
\end{tabular}
\end{table}

\section{ The Computational Method}

In order to find the electronic dipole moments of the nano flakes we performed DFT
calculations on GNFs. We employed density functional theory as
implemented in GAUSSIAN (G09)~\cite{gaussian} which is an
 electronic-structure package that uses a basis set of Gaussian type of orbitals. For the
 exchange-correlation (XC) functional, the hybrid B3LYP~\cite{Becke} is adopted in G09, which
 was shown to give a good representation of the electronic structure in C-based nanoscale systems~\cite{Rudberg}.
 %The self-consistent loop iterates until the change in the total energy is less than $10^{-7}$ eV, and
% the geometries are considered relaxed once the force on each nucleus is less that 50 meV/$\AA$.
 Using the polarized basis set 6-311G** in G09, we expect that our calculation is capable to provide a reliable
 description of the electronic properties of the different systems.
% \textbf{We also performed calculations with
% polarized basis set 6-311G**. The results are converged for polarized basis set and slightly different in energy and dipole
% moment from non-polarized one.}
 Notice that for the H-passivated (F-passivated) GNFs  if the total
 number of electrons or equivalently  $N_H$ e.g.  C$_{21}$X$_{11}$ ($N_F=11$)
 is an odd number the total spin is non zero and therefore in that case  we
  performed spin polarized calculations.

\begin{figure}
%\begin{center}
\centering
\includegraphics[width=9cm,height=8cm]{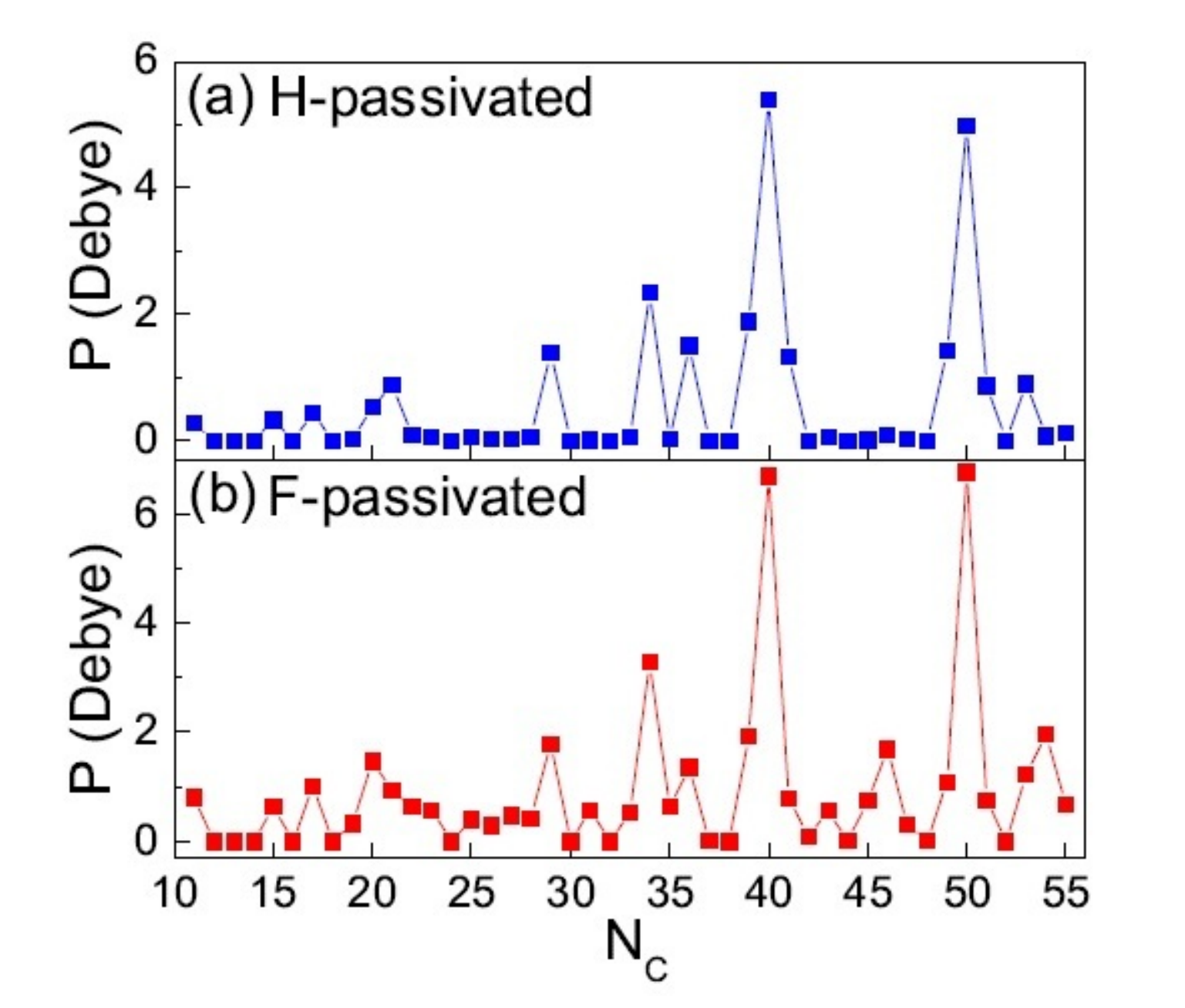}
\caption{(Color online) The absolute value of the  dipole moment
versus the number of carbon atoms for the H-passivated (a) and the
F-passivated (b) GNFs. The largest dipoles are for GNFs with
$N_c=40$ and $N_c=50$ in both cases.\label{figp}}
%\end{center}
\end{figure}

\begin{figure}
%\begin{center}
\centering
\includegraphics[width=0.90\linewidth]{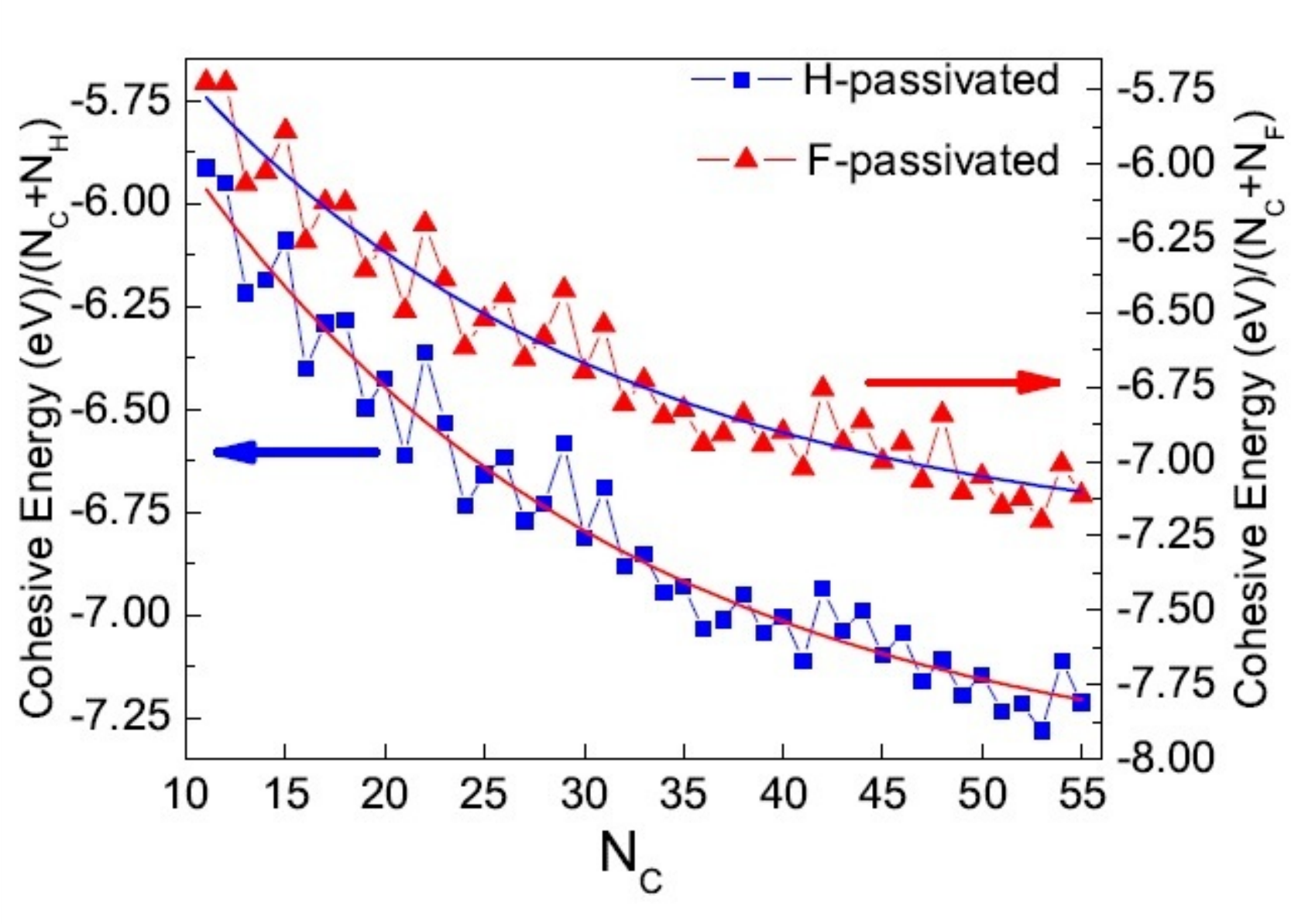}
\vspace{-0.6cm}
\caption{(Color online) The cohesive energy  versus the number of
carbon atoms in H-passivated  and F-passivated  GNFs. The
solid lines are the best fits using to Eq.(~\ref{fitting}).
\label{figE}}
%\end{center}
\end{figure}

\section{Symmetry effect}

In the case of H-passivation (F-passivation), the passivated C
atoms absorb (gives) part of  the electron of the H (to F) atom making the H (F) atom
positively (negatively) charged. This makes the GNFs polarized
with a permanent polarization where the net dipole moment is determined from a
summation over  all local dipoles at the edges.
Therefore, the net dipole moment will depend on the geometry of the system which
we will study in this section. The
electronic and magnetic properties of GNFs originate from this
un-balance in charge distribution at the edges.

In Fig.~\ref{figmodel} we depict the minimum energy configurations
which are saturated by hydrogen or fluorine atoms. These
configurations were obtained using the conjugate gradient
minimization method as outlined  in our previous work~\cite{kosi1}
and revisited in the current study by DFT optimization.
In order to check the minimum energy configuration we performed extra calculations for three typical
isomers with N$_C$=16 (see Fig.~\ref{isomers}). The corresponding binding energy are listed in Table~I. The lowest energy configuration is the 16$_{10}$ structure in Fig.~1.
 The
defective H- and F-passivated small clusters are slightly  buckled
after relaxation.
%The energy of the buckled GNFs is negligible as
%compared to that of planar optimized GNFs using the same functional,
%e.g. for C$_{29}$H$_{15}$ the difference in energy is 0.002 a.u. $(0.2\%)$.
%Therefore, in our study we considered only planar H- and
%F-passivated GNFs.
 We categorized the studied GNFs into two
different groups: i) fourteen GNFs with $n$-fold symmetry, e.g.
C$_{12}$H$_{10}$ with $n=2$, C$_{13}$H$_{9}$ with $n=3$ and so on,
which are shown in Fig.~1(a); and ii) the systems with pentagon and
heptagon defects (shaded) and those without $n$-fold symmetry, e.g.
C$_{11}$H$_{9}$ which are shown in Fig.~1(b). The latter systems may
have  mirror symmetry. Here we limit our study to those C$_{N_c}$
X$_{N_x}$ structures that have minimum energy for planar flakes. The
first aforementioned group has always zero dipole moment while the
second group has non-zero total dipole moment and in some cases are
even a giant polar
 molecule. Notice that larger flakes with n-fold symmetry (the first group) should also have zero total dipole
 moment.

\begin{figure}
%\begin{center}
\centering
\includegraphics[width=0.80\linewidth]{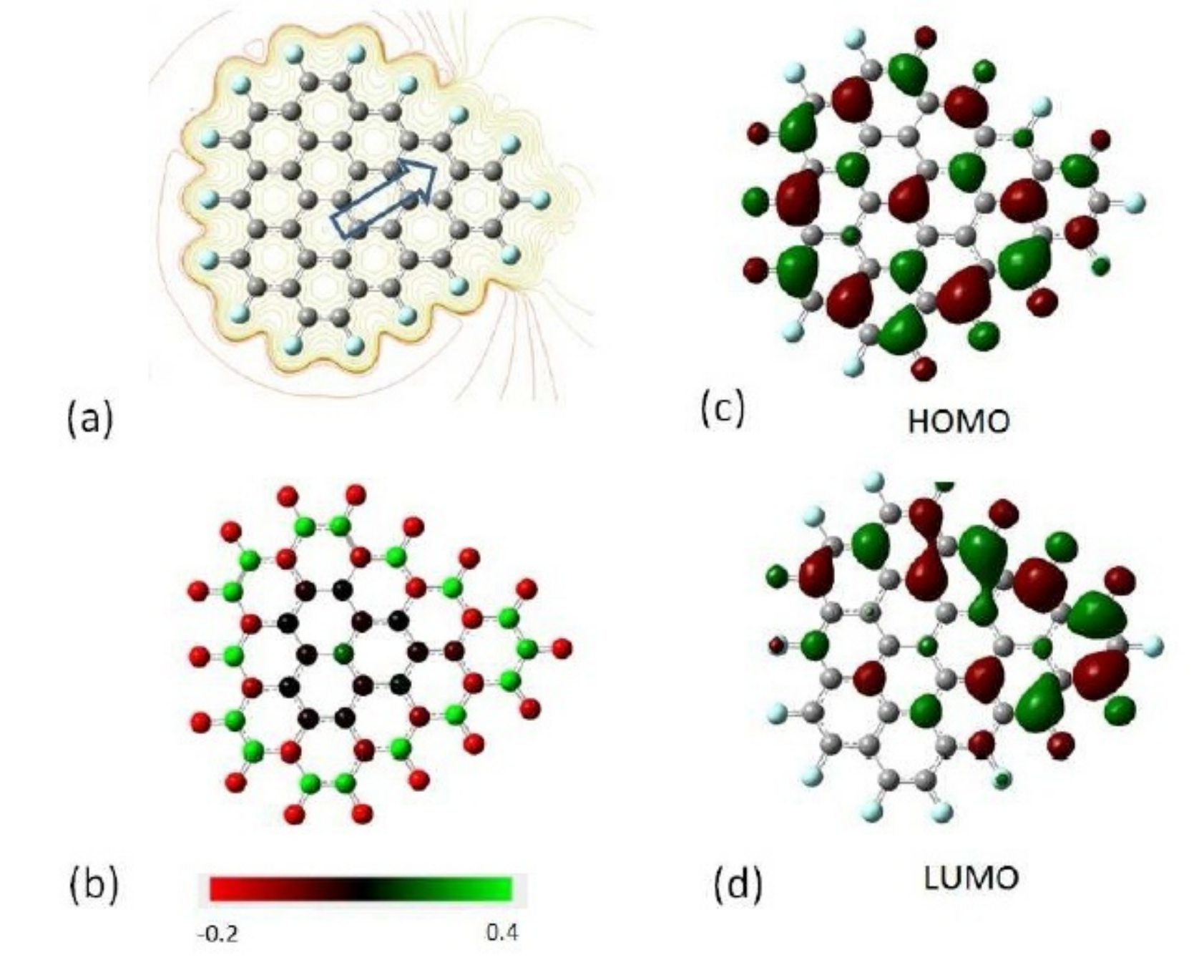}
 \caption{(Color online) (a) Electrostatic potential
contours around F-passivated GNFs - C$_{40}$F$_{16}$- which has the
largest dipole moment indicated by the arrow, (b) Corresponding  charge distribution, (c)  HOMO and (d) LUMO orbitals. The
red and green colors stand for the positive and negative  signs of the
molecular orbital wave function,
respectively. \label{PF40}}
%\end{center}
\end{figure}

 In the first group (Fig.~\ref{figmodel}(a)), e.g. in the 2-fold symmetry cases a rotation of $180^{\circ}$ around the z-axis transforms
$\textbf{r}_i\rightarrow$  $-\textbf{r}_i$ and consequently
$\overrightarrow{P}_T$=0. In the second
group there is no n-fold symmetry where different edges have
different orientation of local dipole moments which do not cancel each
other hence resulting in a non-zero net dipole moment. More defects that are randomly
distributed increase the dipole moment strength.

In Fig.~\ref{figp} we show the absolute value of the total dipole moment
versus $N_c$ for all studied GNFs. In Fig.~\ref{figp}(a) the
H-passivated system and in Fig.~\ref{figp}(b) the F-passivated system are
shown. We see that in both cases  the above mentioned
symmetry issues are obeyed. The net dipole of GNFs with F-passivation are
larger than those for H-passivation which is due to the larger electronegativity
of F. It is interesting to note that the two systems with mirror
symmetry, i.e. C$_{40}$H$_{16}$ and C$_{45}$H$_{17}$ have respectively the largest
- 6.35 Debye - and the  smallest - 0.015 Debye - dipole moment.
 The corresponding dipole moment for C$_{40}$F$_{16}$ and
C$_{45}$F$_{17}$ are 7.69 Debye and 0.61 Debye, respectively. Notice
that the larger component of the dipole moment  is along the symmetry axis. The
system C$_{50}$X$_{18}$ without particular symmetry has the second
largest dipole moment. The ratio between the dipole moments of C$_{40}$F$_{16}$ and
C$_{40}$H$_{16}$ is 1.21.

It is surprising that a simple linear summation over all
local dipole moments in e.g. C$_{40}$X$_{16}$, apparently leads to a zero
dipole moment if one assumes equal $\overrightarrow{p}_{C-X}$ for all
saturated bonds, while this cluster has the
largest dipole moment. The reason for such effect is that
the local dipole moments $\overrightarrow{p}_{C-X}$ are not in the same direction due to
the non-uniform distribution of the C-X second neighbors, i.e. most of the C
atoms in the bottom of C$_{40}$X$_{16}$ (see Fig.~\ref{figmodel}(b)) are
connected to the inner C atoms but at the top of the system there is no
inner C atom to give more electron charge to the corresponding C-X bonds.

\section{Energy difference between HOMO and LUMO}

We study small carbon flakes that are saturated by H or F which have
discrete energy levels and are filled by electrons. In the case of
spin polarized calculations there is an extra energy level. In
Fig.~\ref{figE} we plot the cohesive energy versus $N_c$ for all
studied GNFs. It is seen that the energy per atom decreases with
$N_c$ in both H-passivated and F-passivated GNFs. Increasing the number of C atoms in GNFs increases the
number of C-C bonds more rapidly than the C-X bonds at the edges. Since
the total cohesive energy is a function of C-C and C-X binding
energy we expect that the total cohesive energy rapidly approaches
 the bulk cohesive energy ($e_0$). On average we found that the
energy decreases according to
\begin{equation}
 E/(N_c+N_X)=e_0+be^{-\lambda N_c}
\label{fitting}
\end{equation}
where $e_0$=-7.4 eV, b=-2.37 eV and $\lambda$=0.045 for H-passivated
and $e_0$=-7.3 eV, b=-2.53 eV and $\lambda$=-0.046 for F-passivated
GNFs which are shown by the solid curve in Fig.~\ref{figE}.
Because the C-F binding energy is larger than C-H in F saturated
GNFs the curve for F-passivated is above the one of H-passivated. For
larger N$_{X}$/N$_{C}$ ratio the total cohesive energy is far from the bulk
energy which is due to the so called, i.e. edge effect. Adding a C atom to a typical GNF
decreases the energy if the new formed GNF becomes more stable
which is mostly the case  for those GNFs without pentagon and
heptagon defects. If the next GNF has pentagon or heptagon defects
the energy is larger, see e.g. $N_c$=11,15,17,20,29,31. We
emphasize that most of the studied GNFs are planar-like structures.  Moreover
the systems with a larger ratio $\frac{N_c-N_x}{N_x}$ have lower
energy which corresponds to higher stability due to the  lower
number of edge atoms. It is seen that $\Delta
E=E_{N_{c+2}}-E_{N_c}=constant$ for some $N_c$ (e.g. between
$N_c=$42 and $N_c=$53) where $\Delta N_H=0$ in H-passivated GNFs.
This is promising for quantum dot design: fixing the number of edge
H-atoms and increasing the number of C atoms decreases the energy of
the system with constant increments.
 The latter effect is similar to
the regular jumps observed in the electrochemical potential in
quantum dots. First we  compare the energy of
two typical GNFs ( with only hexagons, e.g. C$_{37}$X$_{15}$) and with
also pentagon and heptagon, e.g. C$_{39}$X$_{15}$) with the same
flakes without X-atoms at the edges (bare flakes). Results show that
in both cases independent of the type of X atom edge passivation
the total energy is reduced, i.e. E(C$_{37}$F$_{15}$)-E(C$_{37}$)=-1.9948 eV,
E(C$_{37}$H$_{15}$)-E(C$_{37}$)=-1.9380 eV, E(C$_{39}$F$_{15}$)-E(C$_{39}$)=-1.9298 eV,
and E(C$_{39}$H$_{15}$)-E(C$_{39}$)=-1.8724 eV. The energy decreases much more
 in C$_{37}$X$_{15}$  than for C$_{39}$X$_{15}$. Therefore the GNFs with regular hexagon edges
becomes more stable when they are saturated by X atoms with respect
to the GNF which contain heptagons and pentagons. This is in
agreement with those reported for triangular GNFs flakes in Ref.~\onlinecite{Voznyy}.

In Fig.~\ref{PF40} we show for C$_{40}$F$_{16}$ (a) the
electrostatic contour, (b) charge distribution,  (c) the
corresponding HOMO and (d) LUMO. The arrow in (a) indicates the
giant permanent dipole in C$_{40}$F$_{16}$. There is a clear
relation between the electrostatic potential and the charge distribution. The LUMO is localized on the parts with transferred
charges to the HOMO region. The less symmetry in any GNFs leads to
the non-horizontal direction of the dipole moment. The dipole moment
is directed to the region with the LUMO orbitals.

In all n-fold symmetric GNFs the net dipole is zero while one
can define local dipoles which eventually cancel each other. Note
that using a different functional for the exchange correlation in our DFT
calculation  may change the energy gap slightly,
however the energy gap for H-passivated and F-passivated GNFs are
close to each other. This is in agreement with the results of
Ref.~[\onlinecite{Zheng2008}].

\begin{figure}
\begin{center}
\includegraphics[width=0.9\linewidth]{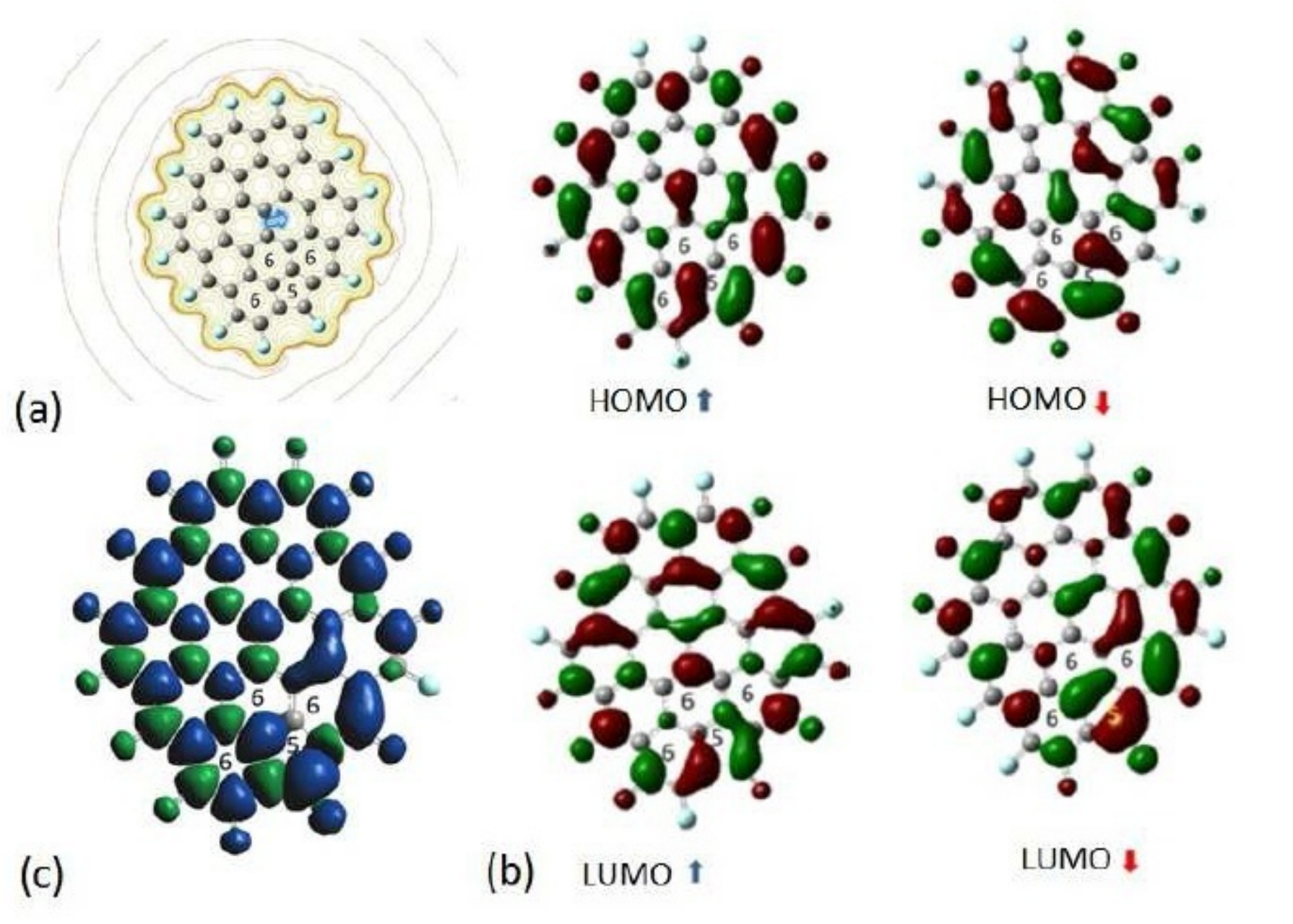}
 \caption{(Color online) Electrostatic potential
contours around C$_{41}$F$_{15}$ with non zero total spin. The
dipole moment is indicated by the arrow. (b) The corresponding spin-up
and spin-down HOMO and LUMO orbitals. The red and green colors stand
for the positive and negative signs of the molecular orbital wave
function, respectively. (c) Isosurfaces of spin density for
C$_{41}$F$_{15}$. Blue and green isosurfaces are 0.0004\,e$/\AA^3$
and -0.0004\,e$/\AA^3$, respectively. \label{PF41}}
\end{center}
\end{figure}

In Fig.~\ref{PF41} we show the electrostatic contour lines around
C$_{41}$F$_{16}$ (a) which has non-zero total spin. The
corresponding spin-up and spin-down HOMO and LUMO are shown in (b).
There is a clear difference in the orientation between orbitals of
spin-up and spin-down which results in a different energy gap
between spin-up and spin-down electrons. The spin-up electrons have
larger energy gap as compared to spin-down. Notice that we do not
show the HOMO and LUMO for all studied GNFs which can be made available upon
request. As a typical case, the isosurface of the spin density
for C$_{41}$F$_{16}$ is shown in Fig.~\ref{PF41}(c). Although the
frontier orbital are mostly nonuniformly distributed at the edges,
the isosurface of the spin density are uniform except on the heptagon's atoms, i.e.
only the $\alpha$ spin has a considerable value different from zero.
In general for a graphene nanoflakes with zigzag edges, the spin magnetizations of A and B sublattices are
aligned antiparallel, i.e. antiferromagnet spin ordering~\cite{28} which is not similar to our studied GNFs.
This alternative is broken by the defect in a way that depend on the out-of-plane distortion~\cite{Akhukov}.

\begin{figure}
\begin{center}
\includegraphics[width=9cm,height=8cm]{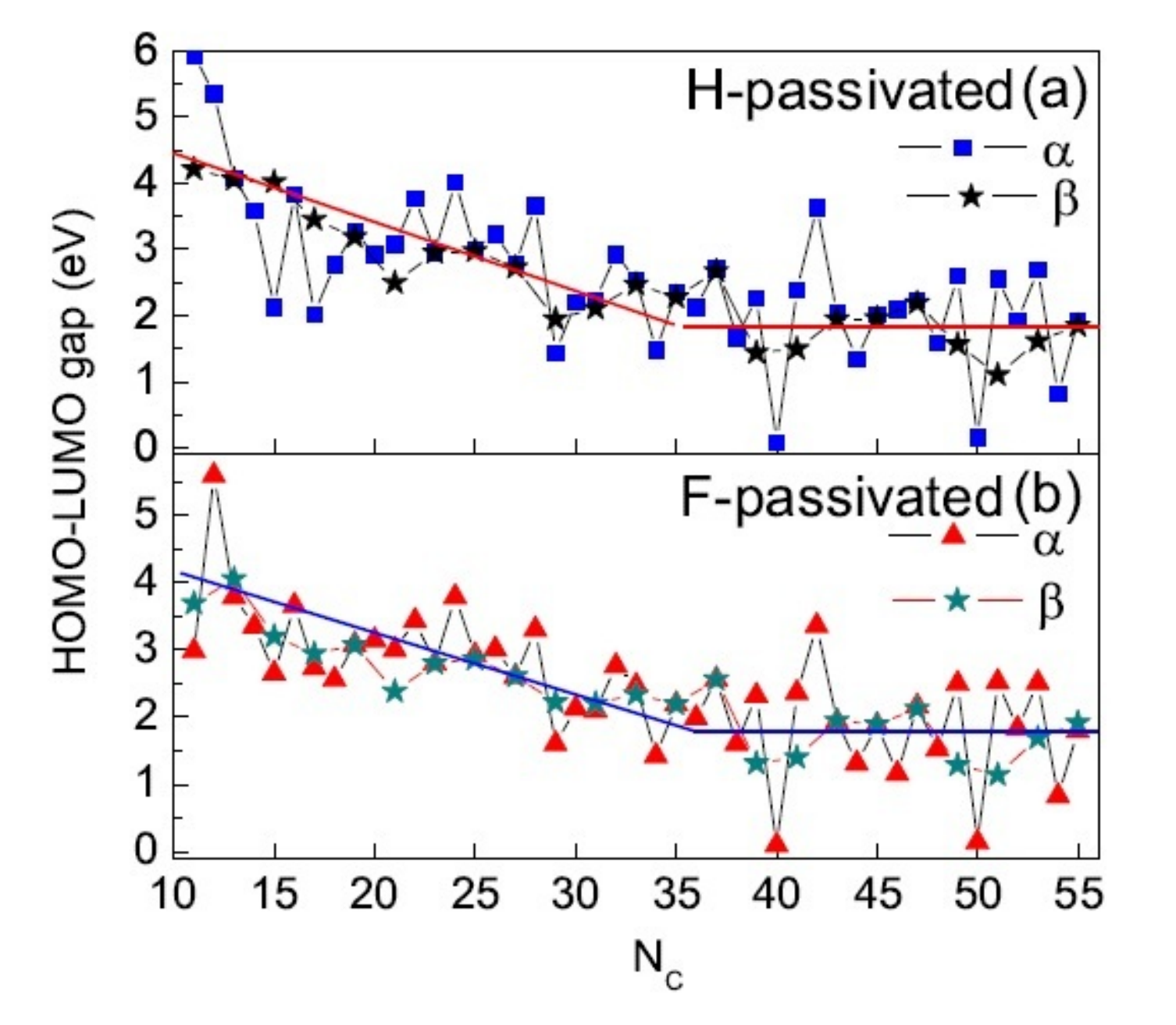}
\caption{(Color online) HOMO$-$LUMO gap versus the number of carbon
atoms in H-passivated (a) and F-passivated GNFs(b). The gap
approaches  zero and is close to zero beyond $N_c=35$. $\alpha$ and
$\beta$ stand for the HOMO$-$LUMO gap for spin up and spin down, respectively.
The solid lines show the overall change in the gaps.
\label{Figgap}}
\end{center}
\end{figure}

In Fig.~\ref{Figgap} we show the HOMO-LUMO energy gap $\Delta$
versus $N_c$ which on average decreases linearly up to $N_c \sim
35$. The larger the electrostatic potential difference between the
ends of GNFs, the lower the energy gap. The electrostatic potential
due to the local charge transfer between C and X decreases the
energy of the LUMO level and increases the energy of the HOMO level
thus reducing the energy gap. We fit a linear line to the results for
both H-passivated and F-passivated systems. This curve is shown in
Fig.~\ref{Figgap}. For large GNFs ($N_c > 35$) the energy gap is
almost constant. It is well known that the energy gap of
graphene is zero. Tight binding calculations predict that in
hexagonal (triangular) GNFs having more than $10^3$ ($10^6$) C atoms
the energy gap approaches  zero~\cite{Guclu2010}. In
 finite size GNFs  the energy gap strongly depends on the
geometry of the system and the type of edges, e.g. zigzag versus armchair.
we found in previous work that the energy gap decreases like a/N where
a=4.9 eV~\cite{ZHANG2008}. On the other hand
experiments on GNFs revealed that the energy gap decreases like 1/L
where L is the lateral dimension of
GNF~\cite{Geim2007, Ritter2009}. Here our
studied GNFs typically have no particular edge structure and we found that the energy gaps decrease up to about
 2 eV.

Although the polarization of GNFs can be understood from the
possible symmetries of GNFs, however the difference between the
energy of the HOMO and the LUMO ($\Delta$) can not be explained
simply as due to an increase of the dipole moment, i.e. the energy
gap is a scaler physical quantity and one does not expect that all
systems with $\overrightarrow{P}_T$=0 have zero energy gap. In other
words there is no straightforward relation between the symmetry and
the energy gap in GNFs. The energy gap is affecten by the
non-uniform distribution of the electrostatic potential which is a
function of the non-uniform charge distribution over the GNFs.
Nevertheless as seen from Figs.~\ref{figp} and ~\ref{Figgap} the GNFs with the
largest dipole moment have the lowest energy gap  hence the polar GNFs
in this study have a lower energy gap which originate from the
un-balanced charge distribution.  This effect strongly depends on
the geometry and can not be generalized to every isomer with the
same number of C atoms.

\begin{figure*}
\begin{center}
\includegraphics[width=0.90\linewidth]{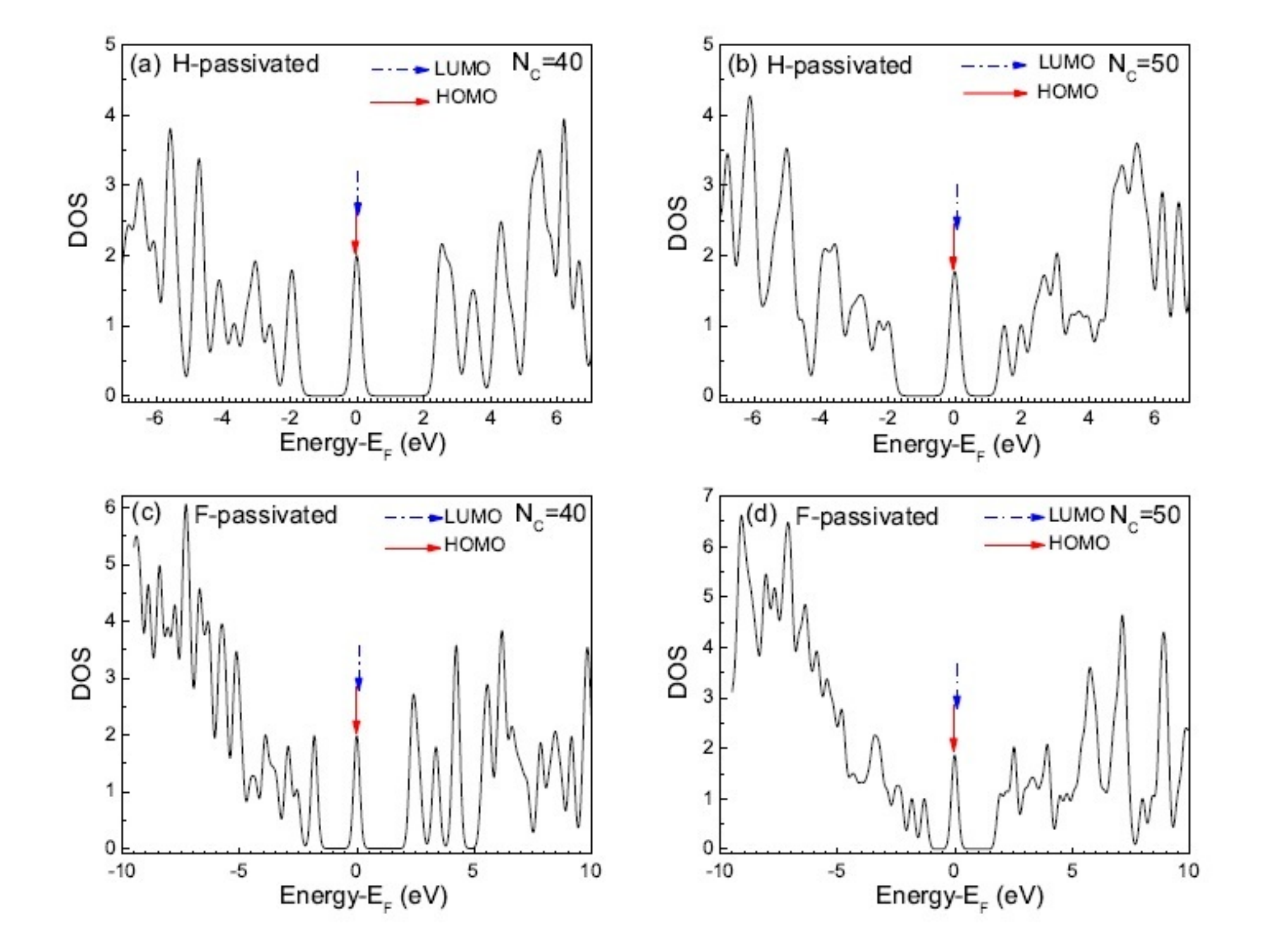}
\caption{(Color online) Density of states  for H-passivated
(a,b) and F-passivated (c,d) graphene nano flakes; the energy
gap in (a,b,c,d) is almost zero while they have maximum dipole moment
among the studied GNFs.  The vertical arrows refer to the position of the HOMO (red) and LUMO (blue).
The Fermi energy was set at zero.
\label{figDos}}
\end{center}
\end{figure*}

\begin{figure*}
\begin{center}
\includegraphics[width=0.9\linewidth]{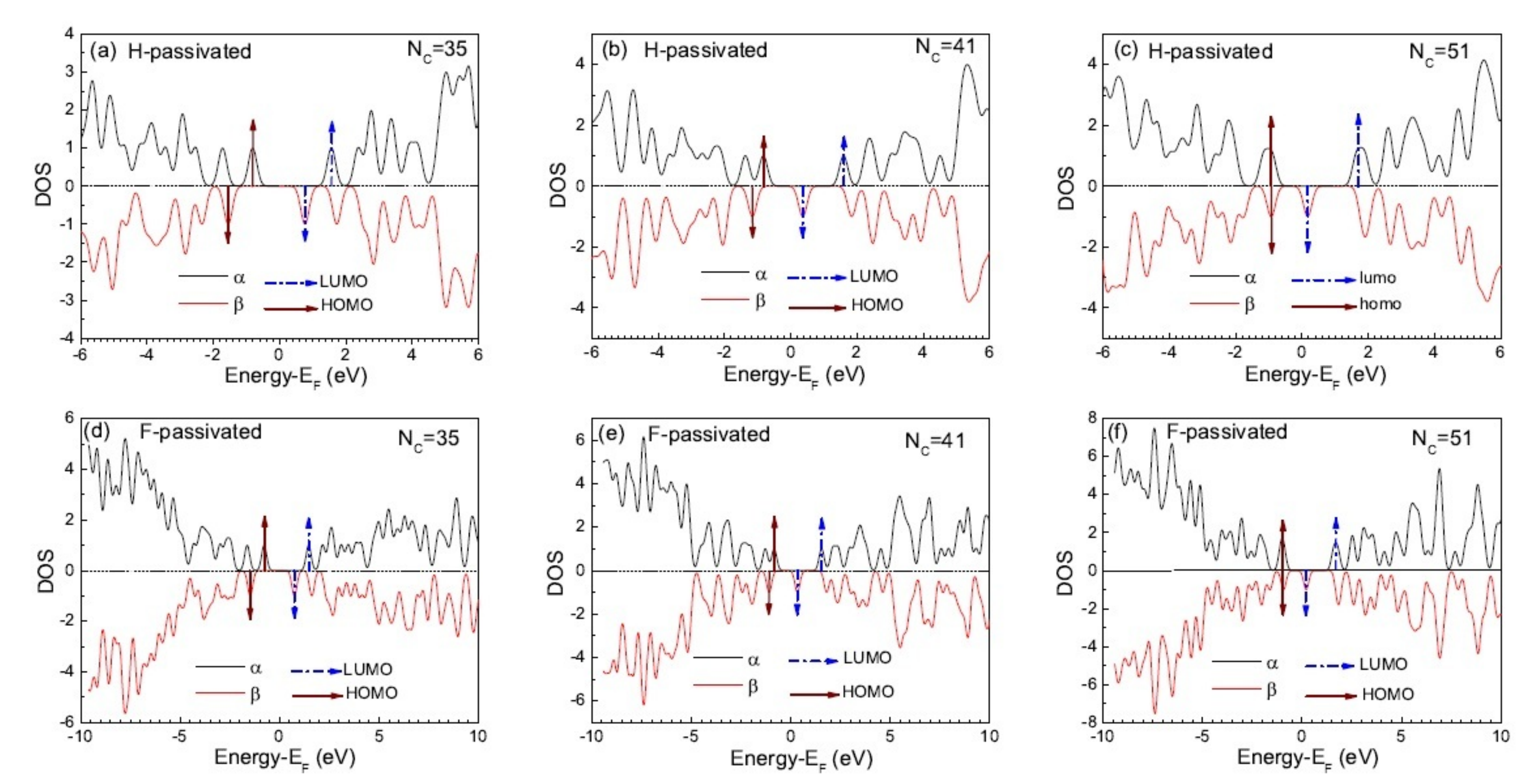}
\caption{(Color online) Density of states  for H-passivated
(a,b,c) and F-passivated (d,e,f) graphene nano flakes with non-zero
total spin; the energy gap in (b,c,e,f) are different for spin-up
(denoted by $\alpha$) and spin-down (denoted by $\beta$)
electrons with a shift in Fermi energy with respect to each other.
In (a,d) the energy gap for spin-up and spin-down are almost the
same with a shift in Fermi energy  with respect to each other. The Fermi
energy occurs at zero.
\label{Dos2}}
\end{center}
\end{figure*}

\begin{figure}
\begin{center}
\includegraphics[width=0.9\linewidth]{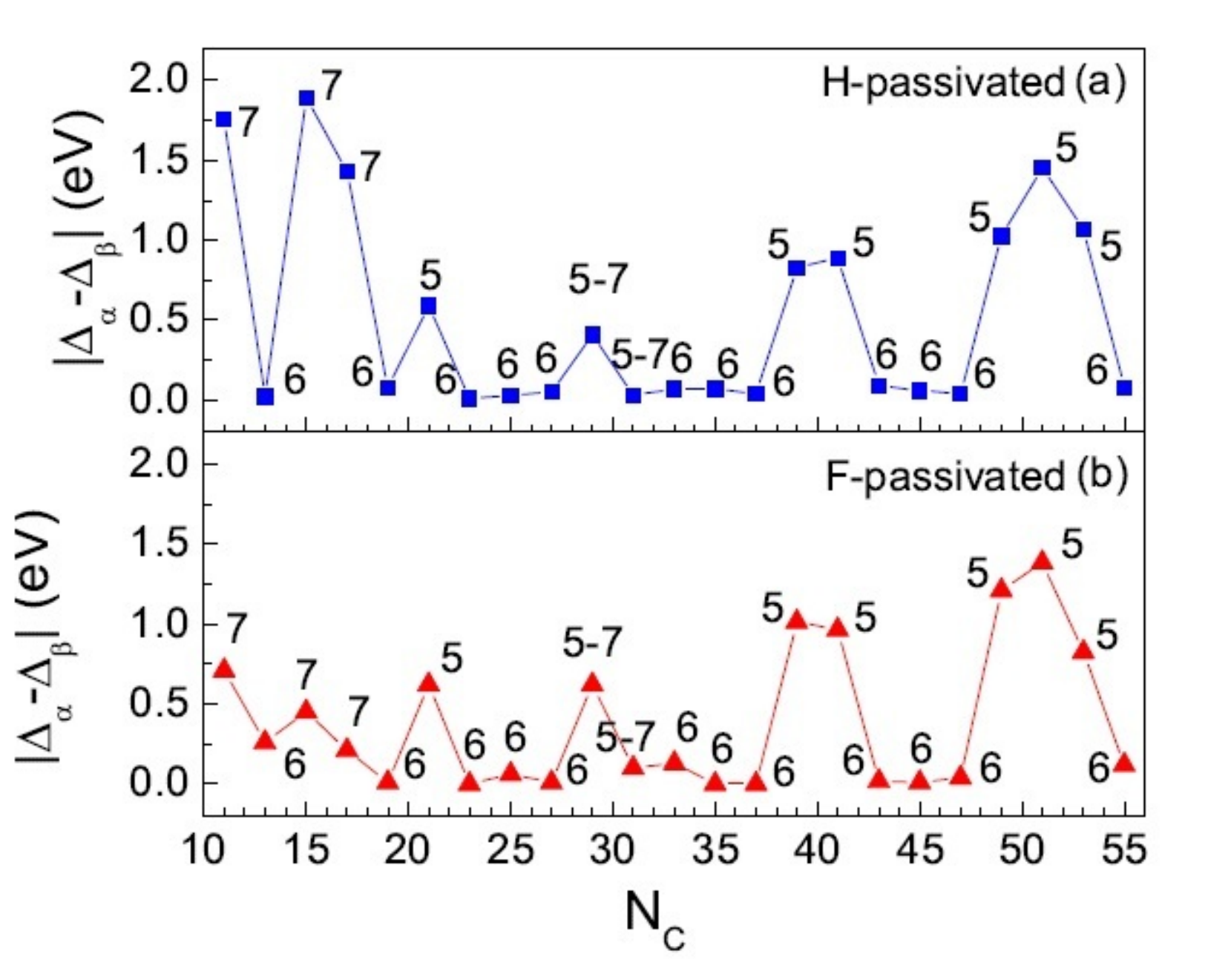}
\caption{(Color online) The energy gap between spin up and spin down for
H-passivated (a) and F-passivated (b) graphene nanoflakes. The numbers refer whether there are
pentagon (5), hexagon (6) or heptagon (7) in the corresponding GNF.
\label{gap_spin}}
\end{center}
\end{figure}

\section{Density of states and spin polarized results}

In Fig.~\ref{figDos} we show the density of states (DOS) for those systems
with the largest dipole moment, i.e. C$_{40}$X$_{16}$ (zero energy gap
$\Delta$) and  C$_{34}$X$_{14}$  with energy gap equal to 1.48 eV for
H-passivated  and 1.42 eV for F-passivated.  We set the Fermi energy at zero
by defining  $E_{F}=(E_{HOMO}+E_{LUMO})/2$. Although the energy gap is almost
zero in GNFs with $N_c=40,50$ but there is a significant gap between
the HOMO, HOMO$_2$ and LUMO and LUMO$_2$ which results in larger
ionization energy for the electrons in HOMO$2$, where index 2 refers to the closest
occupied state to HOMO with larger energy than HOMO. Furthermore the
occupied states in F-passivated GNFs have larger DOS as compared to
H-passivated GNFs which is attributed to the larger charge transfer to
the system by F atoms than H atoms.

 It is also interesting to
investigate the DOS of systems with non-zero total spin. In
Fig.~\ref{Dos2} the DOS of $\alpha$  (spin-up: top panels with black
color) and $\beta$ (spin-down: bottom panels with red color) spins
are shown for three typical systems, i.e. C$_{51}$X$_{17}$,
C$_{41}$X$_{15}$ and C$_{35}$X$_{15}$. The spin-up and spin-down DOS
are almost symmetrical except around the gap region. There is a
clear difference between Figs.~\ref{Dos2}(a) and ~\ref{Dos2}(b)
(and also (c)) while the number of X atoms are the same, $N_x=15$.
In order to  find the Fermi energy we sorted all the energy levels for
spin-up and spin-down and we found the middle point of the new HOMO and LUMO. We found that the
energy gaps between the spin-up and spin-down electrons are
different. In Fig.~\ref{gap_spin} we show the absolute value of the
difference between the energy gap of $\alpha$ spin and $\beta$ spin,
i.e. $\Delta_\alpha-\Delta_\beta$. It is surprising that the
presence of pentagons in GNFs with non-zero total spin maximize
$\Delta_\alpha-\Delta_\beta$. The largest difference is for $N_c=15$
and $N_c=51$ which have one heptagon and one pentagon, respectively.
The symbols `5', `6' and `7' in Figs.~\ref{gap_spin}(a,b) indicate the
presence of pentagon, only hexagons and heptagon in the
corresponding GNFs. Notice that the H-passivated GNFs have a larger
difference between the energy gap of spin-up and spin-down electrons
however the overall pattern in Fig.~\ref{gap_spin} is the same.
Using these results one can predict that for large GNRs with a few
pentagons and$/$or heptagons  half-metallicity can be found.

The GNRs exhibit  half-metallicity  in the presence of
electric field~\cite{son}. It was found that the half-metalicity in
GNRs originates from the fact that the applied electric field
induces an  energy level shift of opposite sign for the spatially
separated spin ordered edge states.

\begin{figure*}
\begin{center}
\includegraphics[width=0.95\linewidth]{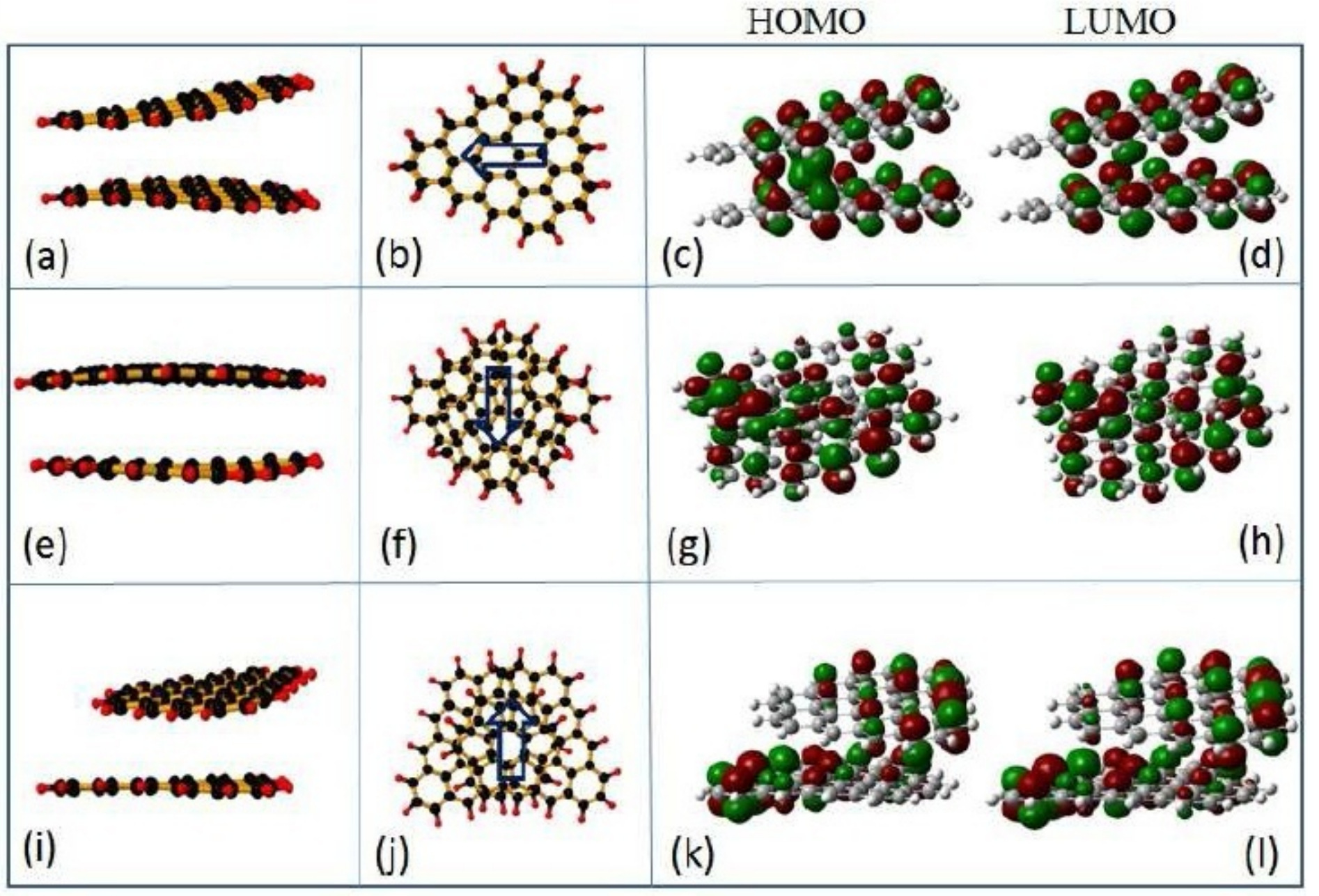}
\caption{(Color online) Bilayer  GNFs C$_{40}$F$_{16}$: parallel with $\theta=0^\circ$(a) side view (b) top view (c) HOMO (d) LUMO;
parallel with $\theta=180^\circ$ (e) side view (f) top view (g) HOMO (h) LUMO; parallel with $\theta=90^\circ$ (i) side view (j) top view (k) HOMO (l) LUMO.
The arrows in Figs.(b,f,j) indicate the direction of polarization.
\label{stacked}}
\end{center}
\end{figure*}

\begin{figure*}
\begin{center}
\includegraphics[width=0.9\linewidth]{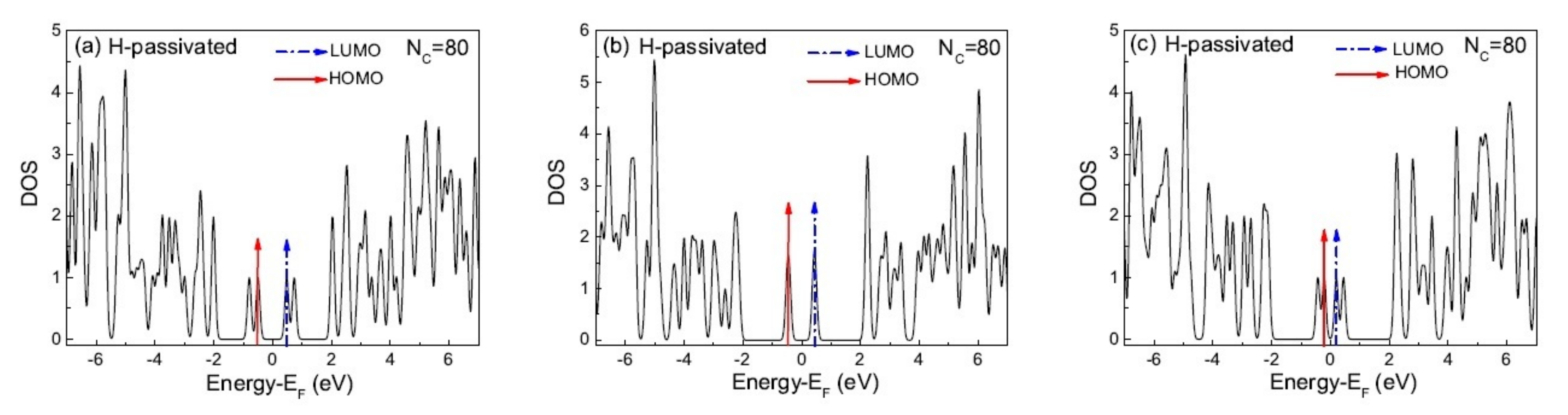}
\caption{(Color online) Density of states  for H-passivated bilayer GNFs (a) parallel with $\theta=0^\circ$ (b) parallel with $\theta=180^\circ$ (c) parallel with $\theta=90^\circ$. An energy
gap appears in contrast to almost zero gap in single GNF C$_{40}$H$_{16}$. The vertical arrows refer to the position of the HOMO (red) and LUMO (blue).
\label{stacked_dos}}
\end{center}
\end{figure*}

\section{Bilayer  GNFs (C$_{40}$H$_{16}$ and C$_{40}$F$_{16}$)}

Bilayer GNFs, which to the best of our
knowledge, have not been investigated from first principles. We select
GNFs which have the largest dipoles. In order to study the bilayer
 GNFs we performed three typical DFT calculations for two bilayer
GNFs C$_{40}$H$_{16}$ and three other DFT calculations
C$_{40}$H$_{16}$ which have different mutual orientation which is
characterized by rotation angle with respect to each other $\theta$,
i.e. parallel with $\theta=0^\circ$,
 (Fig.~\ref{stacked})(a)), parallel with $\theta=180^\circ$
 (Fig.~\ref{stacked}(e))
and parallel with $\theta=90^\circ$ (Fig.~\ref{stacked}(i)).
%During examination of GNFs stacking, we found that the polarity of
%GNFs affects their arrangement.
For the case  $\theta=0^\circ,180^\circ$ GNFs in adjacent layers are arranged
head-to-head (and tail-to-tail) and head-to-tail arrangement,
respectively.

\emph{i)} $\theta=0^\circ$: In Figs.~\ref{stacked}(a)(b) we show a side
and top view, respectively, of the optimized structure for two
bilayer C$_{40}$H$_{16}$ with $\theta=0^\circ$ which after optimization are found to be bent
with positive curvature. GNFs of C$_{40}$H$_{16}$
in the adjacent layers are directed head-to-head and tail-to-tail.
The closest and longest distance between two adjacent layers (not
including the hydrogen atoms) are found to be 3.27 and 4.83\,\AA,~
respectively, which are closer and longer than the distance between
graphite layers, i.e. 3.35\,\AA~. The optimized structure indicates
charge repulsion in the ends (tails) in bilayer GNFs. Because
C$_{40}$H$_{16}$ is a giant polar molecule, we can understand the
repulsion of two bilayer molecules as the dipole-dipole interaction,
i.e. $U=\vec{P}_1.\vec{P}_2/R^3$ where $R$ is the perpendicular distance
between two dipoles. If the
sheets do not bent we expect that $\theta=180^\circ$ has lower energy
than $\theta=0^\circ$, however bending reduces the energy and makes both
energies close to each other. In Table II we listed all results for
bilayer C$_{40}$H$_{16}$ and C$_{40}$F$_{16}$.

The HOMO-LUMO and DOS for $\theta=0^\circ$ are illustrated in
Figs.~\ref{stacked}(c,d) and Figs.~\ref{stacked_dos}(a),
respectively. There is a clear mixing of orbitals between two HOMOs
in adjacent GNFs (see Figs.~\ref{stacked}(c)). This is different
from the stacking of graphite where the layers are parallel and there is no
electron-sharing between two layers and the interaction is mostly
a weak vander Waals interaction. This mixing effect reduces the energy,
otherwise we expect that the $\theta=0^\circ$ case has higher energy
than $\theta=180^\circ$. The HOMO-LUMO gap 0.97 $eV$ appears in bilayer
GNFs which was found to be almost zero for a single GNF
C$_{40}$H$_{16}$, see Fig.~\ref{Figgap}. Now there is a small gap
between the HOMO, HOMO$_2$ and LUMO and LUMO$_2$ in contrast to
single GNF C$_{40}$H$_{16}$.

\emph{ii)} $\theta=180^\circ$: For this case the head of the top GNF is
directed towards the tail of the bottom GNF. The optimized structure
leads to a relative rotation, i.e. the optimized
structure does not satisfy the condition $\theta=180^\circ$. Here
curvature is negative and the closest distance between different
layers (not including hydrogen atoms) thus appears in the vicinity of
these head-tail, and the distance of the head1-tail2 and tail1-head2
between two layers are 3.46 and 4.02 $\AA$, respectively. Therefore
the distances are larger than in graphite stacking and we
expect a weak interaction between the two GNFs. By careful examination of
bilayer GNFs, we do not found that the interatomic distance in different
molecules is closer than this value. The HOMO-LUMO and DOS are shown
in Figs.~\ref{stacked}(g,h) and Figs.~\ref{stacked_dos}(b),
respectively. There is no mixing in the frontier orbital. The energy
is slightly higher than in the case with $\theta=0^\circ$. Note that if the
GNFs does not bent the mutual interaction energy of $\theta=180^\circ$
could be lower than $\theta=0^\circ$. The energy gap is found to be 0.82
$eV$ which is less than for  the $\theta=0^\circ$ case. This is due to the fact
that the bilayer GNFs with $\theta=180^\circ$ has a larger net dipole and
a different orientation as compared to the previous case.

\emph{iii)} $\theta=90^\circ$: The tail of the top GNF is directed towards
the head and away from the tail of the bottom GNF. After optimization the
mutual angle is not $\theta=90^\circ$ and there is an additional mutual
rotation. The closest and longest distances between different layers
(not including hydrogen atoms) thus appear in the vicinity of these
head-tail and tail-tail, are 3.35 and 4.89 $\AA$, respectively,
which are equal and longer than the graphite stacking distance,
respectively. The HOMO-LUMO and DOS are shown in
Figs.~\ref{stacked}(k,l) and Figs.~\ref{stacked_dos}(c),
respectively. There is no mixed frontier orbitals and the energy is
larger than those for $\theta=0^\circ$ and $\theta=180^\circ$. The net dipole
is much larger than those of $\theta=0^\circ$ and $\theta=180^\circ$ resulting
to a lower energy gap, i.e. 0.41 $eV$. Note that for two GNFs
consisting of two C$_{40}$F$_{16}$ our calculation was not fully
optimized, hence $\theta=180^\circ$ is not likely the preferential bilayer structure for
this GNFs.

Based on the above results, we may reasonably regard that for a
large-polar GNFs such as C$_{40}$H$_{16}$, the polarity of GNFs
determines the mutual orientation of the GNFs arrangement in the
crystal structure. Additionally, it is interesting to note that
while a single C$_{40}$H$_{16}$ is planer, however it bends when it
interact with other GNFs which is due to its finite size (the edges
interact). The resulting  bilayer structures depend on the type of stacking and  there is a tilting of the flakes.

\begin{table}[tp]%
\caption{Net dipole moment, energy gap and cohesive energy for bilayer
GNFs with different mutual orientation}
\begin{tabular}{c  | c  l  l}
\hline
  & P(Debye)     & Gap (eV)      & $\frac{Energy}{N_C+N_{H,F}}$(eV/atom)            \\

 \hline
$\theta=0^\circ$       &  $$                         &   $$             &$$ \\
C$_{40}$H$_{16}$  & $0.006$                      & 0.99        & -7.012\\
C$_{40}$F$_{16}$  & 0.781                      & 0.90           & -6.905\\
 \hline
$\theta=180^\circ$ &                            &               & \\
C$_{40}$H$_{16}$  & 0.059                      & 0.82           & -7.010\\
C$_{40}$F$_{16}$   & -                      & -           &- \\
 \hline
$\theta=90^\circ$ &                           &                & \\
C$_{40}$H$_{16}$   & 0.999                      & 0.41           & -7.008\\
C$_{40}$F$_{16}$   & 1.977                      & 0.38           & -6.904\\
 \hline
\end{tabular}
\end{table}

\section{Raman Spectroscopy}
Raman spectroscopy~\cite{Ferrari2013} is a non-destructive
and quick characterization technique which gives structural and
electronic information. For graphene nano-flakes, the most intense
peaks are G and D, appearing around 1585 cm$^{-1}$ and 1350
cm$^{-1}$, respectively. The G peak corresponds to the C-C bond
stretching, i.e. the first order Raman-allowed E$_{2g}$ phonon at
the Brillouin zone center. The D peak requires a defect to be
activated. Density functional calculations using the hybrid functional
B3LYP with polarized basis set 6-311g** have been carried out in order to obtain the  Raman
spectrum (we used standard method for scaling of  the
frequencies which has been adopted in GAUSSIAN09 software).

\begin{figure}
\begin{center}
\includegraphics[width=0.9\linewidth]{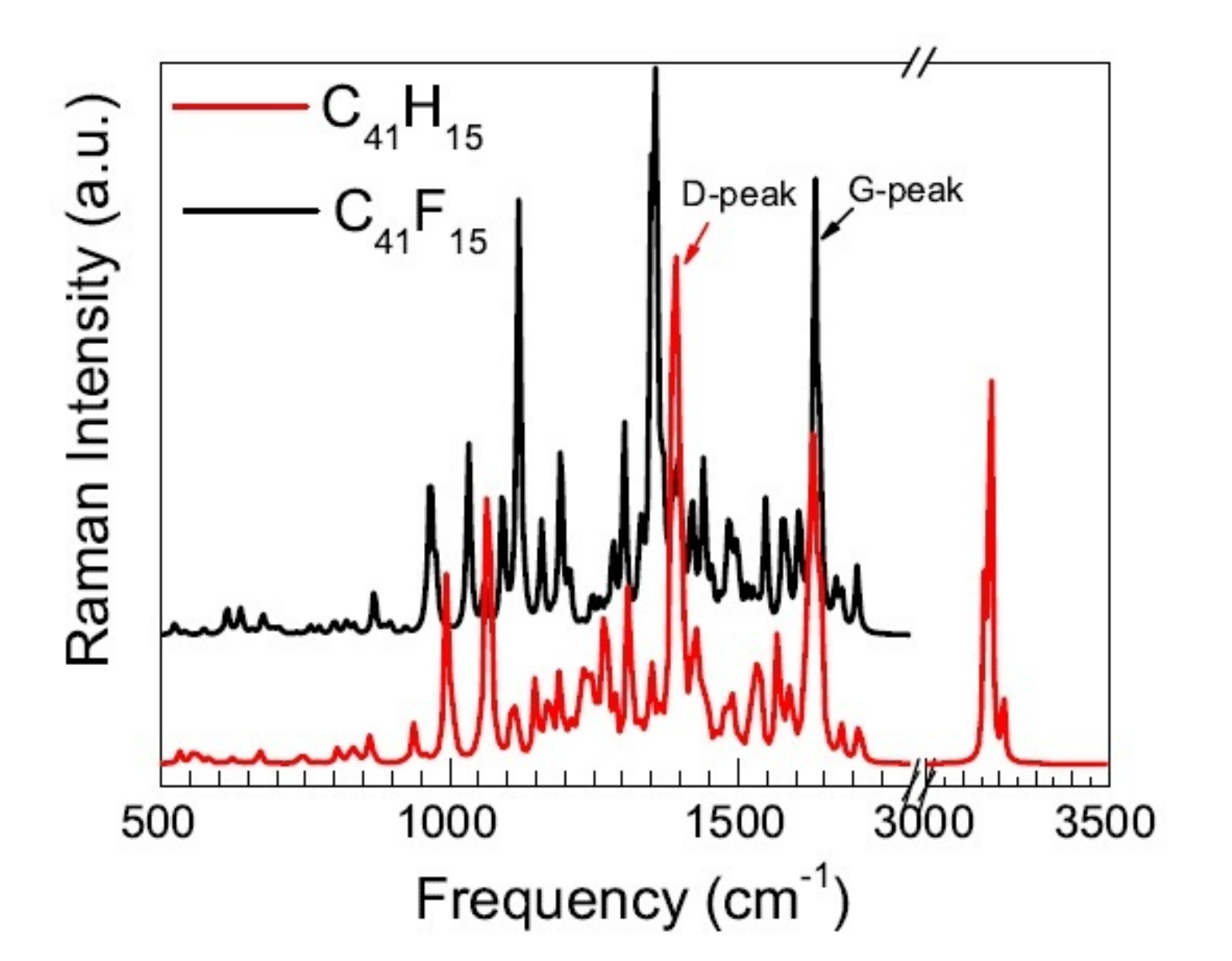}
\vspace{-1cm}
\caption{(Color online) Raman spectrum of  C$_{41}$H$_{15}$ and
C$_{41}$F$_{15}$ nano-flakes.
\label{raman_F}}
\end{center}
\end{figure}

\begin{figure}
\begin{center}
\includegraphics[width=0.9\linewidth]{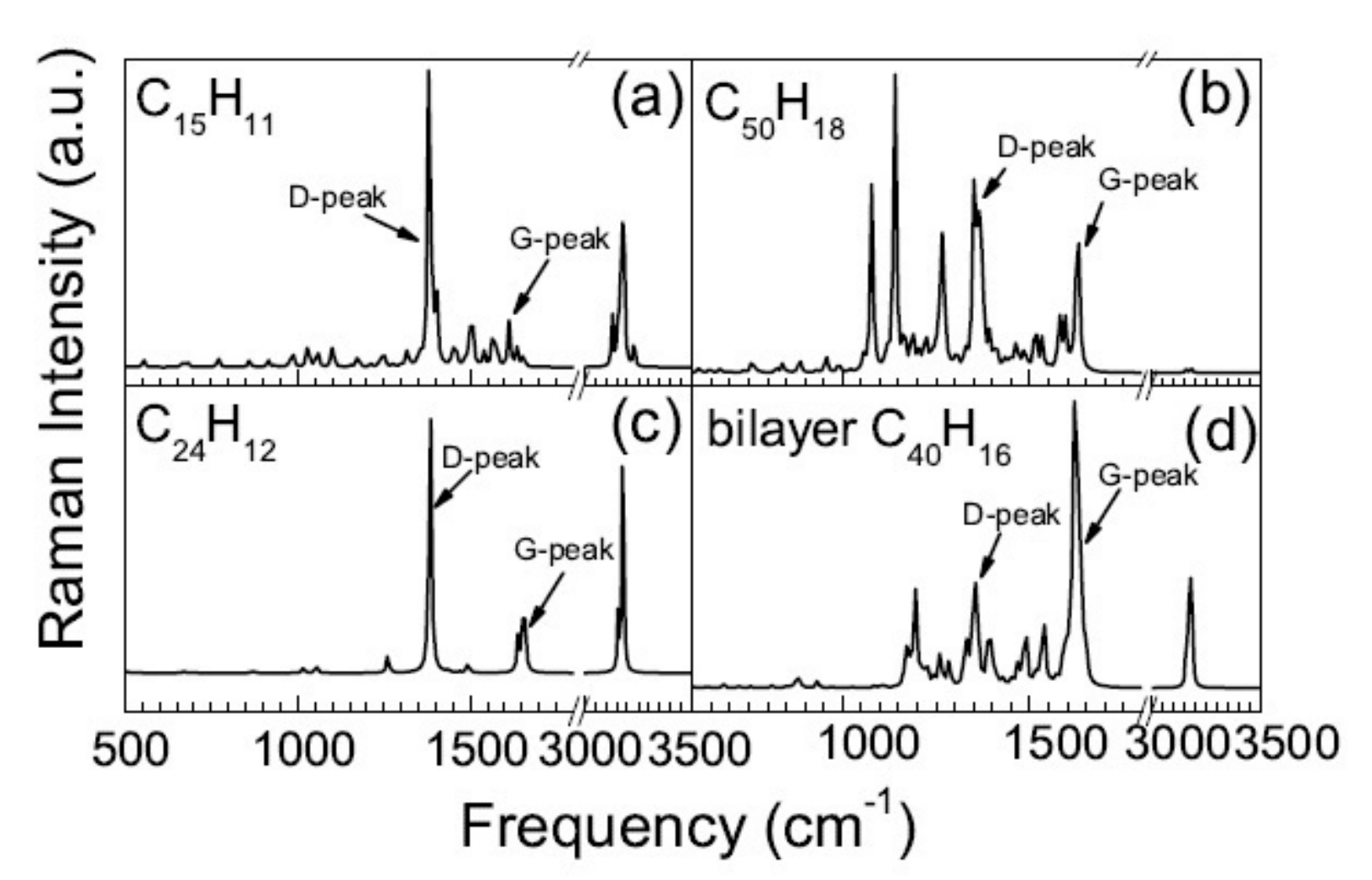}
\vspace{-2.25cm}
\caption{(Color online) Raman spectrum  of H-passivated nano-flakes: (a) C$_{15}$H$_{11}$, (b) C$_{50}$H$_{18}$, (c) C$_{24}$H$_{12}$, and
(d) bilayer C$_{40}$H$_{16}$.
\label{raman_H}}
\end{center}
\end{figure}

Raman spectra are  calculated for some typical H- and F-
passivated nano-flakes (as shown in Fig.~\ref{figmodel}) and bilayer
GNFs. We notice that the introduction of a pentagon in the
C$_{41}$H$_{15}$ nano-flake introduces more high frequency modes
beyond the G-peak as compared to non-defective nano-flake (e.g.
C$_{50}$H$_{18}$ as shown in Fig.~\ref{raman_H} (b)), which is due
to the fact that modes localize around the defect (as shown in
Fig.~\ref{raman_F}). Introduction of a pentagon in the
C$_{41}$F$_{15}$ also introduces more high frequency modes similar
to H-pasivated clusters. Furthermore, a prominent feature at 1260
cm$^{-1}$ characteristic of covalent C-F bond stretching~\cite{Bon}
has been observed. D peak can be found in all GNFs due to the finite
crystalline size where the  edges of the nano-flakes can be seen as defects.
The ratio of the intensities of  D and G peak increases as the
nano-flake size decreases due to the relative increase of the number
of edge atoms and  the less ordered crystalline structure in smaller GNFs (see Fig.~\ref{raman_H}(a)).
For highly symmetric nano-flakes, e.g. C$_{24}$H$_{12}$  only G- and
D-peaks are dominant.  A localized mode is found to
appear at $\thickapprox$\,3100 cm$^{-1}$ in H-passivated GNFs, which
is the typical vibrational mode of the C-H bond~\cite{Demaison}.

For a parallel bilayer GNFs, broad G- and D- bands appear
(Fig.~\ref{raman_H}(d)).
We also found that the intensity of the Raman spectra in GNFs are different than those experimentally
reported for graphene~\cite{Ferrari2013}. Firstly, because of the edge effects which causes non-uniform
distribution of C-C bonds over the GNFs and secondly because of the large ratio between the number of defects
and the total number of carbon atoms in each particular GNF. These enhance the  intensity of D peak with
respect to G peak while they are not shifted with respect to graphene.
\section{Conclusion}

Using extensive ab-initio calculations we studied the electronic
properties of GNFs with several different number of carbon atoms and
two different atoms for edge termination. The n-fold symmetry causes
no net dipole in GNFs. Breaking the n-fold symmetry by heptagon and
pentagon defects and reducing the symmetries to mirror symmetry
enhances the polarization. We found that the larger the dipole
moment  the lower the energy gap for both type of saturated atoms.
The cohesive energy of GNFs reduces with increasing carbon atoms for
constant number of passived  atoms. On average the  energy gap
decreases rapidly. The electrostatic potential around GNFs control
both the polarization and the energy gap of the GNFs. Our spin
polarized calculations show that the difference between the energy
gap of up and down spin is maximized mostly for  GNFs with pentagon
(and heptagon) defects. The H-passivated GNFs have a larger
difference between the spin-up and spin-down energy gap as compared
to F-passivated GNFs. The bilayer GNFs reveal a clear dipole-dipole
interaction which is a consequence of the mutual orientation between
the permanent dipoles in the system. The bilayer GNFs are not
necessarily planar structures and may have a curved structure.

{\emph{ACKNOWLEDGMENTS}} This work was supported by the EU-Marie
Curie IIF postdoc Fellowship/299855 (for M.N.-A.), the
ESF-EuroGRAPHENE project CONGRAN, the Flemish Science Foundation
(FWO-Vl), and the Methusalem Foundation of the Flemish Government.

\end{document}